\documentstyle[preprint,aps,prl]{revtex}

\tightenlines %% LANL requests single line spacing to save paper.

\preprint{
\hbox to \hsize{
\hfill$\vcenter{\hbox{\bf OKHEP-01-03R}
                \hbox{\bf hep-ph/0105250}
                \hbox{July 2001}}$ }
}

\begin{document}
%\draft

%-----------------------------------
% Title
%-----------------------------------
\title{Constraints on extra dimensions from cosmological and terrestrial
measurements}

%-----------------------------------
% Authors
%-----------------------------------
\author{Kimball A. Milton,\thanks{Email: milton@physics.ou.edu}
 Ronald Kantowski,\thanks{Email: ski@physics.ou.edu} 
 Chung Kao,\thanks{Email: kao@physics.ou.edu} and 
 Yun Wang\thanks{Email: wang@physics.ou.edu}}
%-----------------------------------
% Address
%-----------------------------------
\address{Department of Physics and Astronomy, The University of Oklahoma,
Norman, OK 73019}

\date{\today}

\maketitle

%-----------------------------------
% Abstract
%-----------------------------------
\begin{abstract}
If quantum fields exist in extra compact dimensions, they will
give rise to a quantum vacuum or Casimir energy.  That vacuum energy will
manifest itself as a cosmological constant.  The fact that supernova
and cosmic microwave background data indicate that the cosmological constant
is of the same order as the critical mass density to close the universe
supplies a lower bound on the size of the extra dimensions.  Recent laboratory
constraints on deviations from Newton's law place an upper limit.  The
allowed region is so small as to suggest that 
either extra compact dimensions do not exist, 
or their properties are about to be tightly constrained by experimental data.
\end{abstract}

%=======================================================================
%   BEGIN MAIN TEXT
%=======================================================================

It has been appreciated for many years that there is an apparently fundamental
conflict between quantum field theory and the smallness of the cosmological
constant \cite{weinberg}.  This is because the zero-point energy of the
quantum fields (including gravity) in the universe should give rise
to an observable cosmological vacuum energy density,
\begin{equation}
u_{\rm cosmo}\sim{1\over L_{\rm Pl}^4},
\end{equation}
where the Planck length is
\begin{equation}
L_{\rm Pl}=\sqrt{G_N}=1.6\times 10^{-33}\,\mbox{cm}.
\end{equation}
(We use natural units with $\hbar=c=1$.  The conversion factor is
$\hbar c \simeq 2 \times 10^{-14}\,\mbox{GeV\,cm}$.)  This means that the
cosmic vacuum energy density would be
\begin{equation}
u_{\rm cosmo}\sim 10^{118} \mbox{ GeV\,cm}^{-3},
\label{ccprob}
\end{equation}
which is 123 orders of magnitude larger than the critical mass density
required to close the universe:
\begin{equation}
\rho_c={3H_0^2\over8\pi G_N}=1.05\times 10^{-5}h_0^2 \,\mbox{GeV\,cm}^{-3},
\end{equation}
in terms of the dimensionless Hubble constant, 
$h_0=H_0/100 \; \mbox{km\,s}^{-1}\mbox{Mpc}^{-1}$. 
~From relativistic covariance 
the cosmological vacuum energy density must be the $00$ component 
of the expectation value of the energy-momentum tensor, 
which we can identify with the cosmological constant:
\begin{equation}
\langle T^{\mu\nu}\rangle =-u g^{\mu\nu}=-{\Lambda\over8\pi G}g^{\mu\nu}.
\end{equation}
[We use the metric with signature $(-1,1,1,1)$.]
Of course this is absurd with $u$ given by Eq.~(\ref{ccprob}), which would
have caused the universe to expand to zero density long ago.

For most of the past century, it was the prejudice of theoreticians that
the cosmological constant was exactly zero, although no one could give
a convincing argument.  Recently, however, with the new data gathered
on the brightness-redshift relation for very distant type Ia supernov\ae\
\cite{reiss,perlmutter}, corroborated by the balloon observations of the
anisotropy in the cosmic microwave background 
\cite{Boomerang,balbi,DASI}, it
seems clear that the cosmological constant is near the critical value,
or $\Omega_\Lambda=\Lambda/8\pi G\rho_c \simeq 0.6-0.7$ \cite{Boomerang,DASI}. 
 It is very hard to understand how the cosmological
constant can be nonzero but small.

We here present a plausible scenario for understanding this puzzle.
It is reasonable (but by no means established) that vacuum fluctuations
in the gravitational and matter fields in flat Minkowski space give
a zero cosmological constant.\footnote{This is in line with
considerations of Casimir energies in other contexts.  For example,
although the electromagnetic Casimir energy of a ball of dilute
nondispersive dielectric material is divergent, that divergence
can be unambiguously removed as an unobservable bulk and surface
effect, and a unique
finite energy, interpretable as a sum of van der Waals energies,
emerges.  See Refs.~\cite{dielectricball}.} (See below.) Effects due to
curvature are negligible.
But since the work of Kaluza and Klein \cite{kk}
it has been an exciting possibility that there exist extra dimensions
beyond those of Minkowski space-time.  
%Why do not we experience those dimensions?  
Why do we not experience those dimensions?  
The simplest possibility seems to be that those extra dimensions
are curled up in a space $\cal S$ of size $a$, smaller than some observable
limit.

Of course, in recent years, the idea of extra dimensions has become
much more compelling.  Superstring theory  requires at least 10 dimensions,
six of which must be compactified, and the putative M theory, supergravity,
is an 11 dimensional theory.  Perhaps, if only gravity experiences the
extra dimensions, they could be of macroscopic size.  Various scenarios
have been suggested \cite{add,lrs}. 

Macroscopic extra dimensions imply deviations from Newton's law at such
a scale.  A year ago, millimeter scale deviations seemed plausible, and
many theorists hoped that the higher-dimensional world was on the brink
of discovery. Experiments were initiated \cite{long}.  Very recently,
the results of the first definitive experiment have appeared 
\cite{hoyle}, which
indicate no deviation from Newton's law down to 218 $\mu$m.  This poses a
serious constraint for model-builders.\footnote{We might also mention
short distance constraints on Yukawa-type corrections to the gravitational
potential coming from Casimir measurements themselves \cite{most}.}

Here we propose that a very tight constraint indeed emerges if we recognize
that compact dimensions of size $a$ necessarily possess a quantum
vacuum or Casimir
energy of order $u(z)\sim a^{-4}$.  That such energies are observable is
confirmed by recent experiments \cite{exp}.
  These can be calculated in simple
cases.  Applequist and Chodos \cite{ac} found that the Casimir energy
for the case of scalar field on a circle, ${\cal S}=S^1$, was
\begin{equation}
u_C=-{3\zeta(5)\over64\pi^6a^4}=-{5.056\times10^{-5}\over a^4},
\end{equation}
which needs only to be multiplied by 5 for graviton fluctuations.
The general case of scalars on ${\cal S}=S^N$, $N$ odd, was considered
by Candelas and Weinberg \cite{cw}, who found that the Casimir energy
was positive for $3\le N\le 19$, with a maximum at $N=13$ of
$u_C=1.374\times 10^{-3}/a^4$.  The even dimensional case was much
more subtle, because it was divergent.  Kantowski and Milton \cite{km}
showed that the coefficient of the logarithmic divergence was unique,
and adopting the Planck length as the natural cutoff, found
\begin{equation}
S^N, \,\,N \mbox{ even}: \quad u^N_C={\alpha_N\over a^4}\ln{a\over L_{\rm Pl}},
\end{equation} but $\alpha_N$ was always negative for scalars.
In a second paper \cite{km2} they extended the analysis to vectors, tensors,
fermions, and to massive particles, among which cases positive values of the
(divergent) Casimir energy could be found.  Some representative results for
massless particles are shown in Table \ref{tab1}. In an unsuccessful 
 attempt to find stable configurations, the analysis was extended
to cases where the internal space was the product of spheres \cite{birm}.

It is important to recognize that these Casimir energies correspond to a
cosmological constant in our $3+1$ dimensional world, not in the extra
compactified dimensions or ``bulk.''  They constitute an effective source
term in the 4-dimensional Einstein equations.  That there is a correlation 
between the currently favored value of the cosmological constant and
submillimeter-sized extra dimensions has been noted qualitatively before
\cite{kaplanwise,chen}.

 The goal, of course, in all these investigations was to include graviton
fluctuations.  However, it immediately became apparent that the results
were gauge- and reparameterization-dependent unless the DeWitt-Vilkovisky
formalism was adopted \cite{devil}.  This was an extraordinarily difficult
task.  Only in the past year has the general analysis for gravity 
appeared,\footnote{A few special cases were known earlier.  Besides that 
of ${\cal S}=S^1$, the general six-dimensional background was considered by 
Cho and Kantowski \cite{ck91}, which includes ${\cal S}=S^1\times S^1$ 
and $S^2$.} with results
for a few special geometries \cite{chokan}. Cho and Kantowski obtain the unique
divergent part of the effective action for ${\cal S}=S^2$, $S^4$, and $S^6$,
as polynomials in $\Lambda a^2$. (Unfortunately, once again, they are unable to 
find any stable configurations.) The results are also shown in Table \ref{tab1},
for $\Lambda a^2\sim G/a^2\ll1$.
It will be noted that graviton fluctuations dominate matter fluctuations,
except in the case of a large number of matter fields in a small number
of dimensions.
 Of course, it would be very interesting to know the
graviton fluctuation results for odd-dimensional
spaces, but that seems to be a more difficult calculation; it is far easier to
compute the divergent part than the finite part, which is all there is in
odd-dimensional spaces.

These generic results may be applied to recent popular scenarios.  For example,
in the ADD scheme \cite{add} only gravity propagates in the bulk, while the
RS approach \cite{lrs} has other bulk fields in a single extra dimension.

Let us now perform some simple
estimates of the cosmological constant in these models.  The data suggest a
positive cosmological constant, so we can exclude those cases where the
Casimir energy is negative.  For the odd $N$ cases, where the Casimir energy
is finite, let us write
\begin{equation}
S^N, \quad N \,\mbox{odd}:\quad
u_C^N={\beta_N\over a^4}, 
\end{equation}
so merely requiring that this be less than the critical density $\rho_c$
implies ($\beta>0$)
\begin{equation}
a\gtrsim\beta^{1/4}h_0^{-1/2} 67\, \mu\mbox{m}\approx \beta^{1/4} 80\,
 \mu\mbox{m},
\label{betalim}
\end{equation}
taking \cite{pdg} $h_0=0.7$ (with about a 10--20\% uncertainty). 
 As seen in Table \ref{tab2} these lower limits (for a single species) are still
an order of magnitude below the experimental upper limit \cite{hoyle}.
Much tighter constraints appear if we use the divergent results for even
dimensions.  We have the inequality ($\alpha>0$)
\begin{equation}
a\gtrsim[\alpha\ln(a/L_{\rm Pl})]^{1/4}80\,\mu\mbox{m},
\end{equation}
where we can approximate $(\ln a/L_{\rm Pl})^{1/4}\approx2.9$.
Again results are shown in Table \ref{tab2},
which rules out all but one of the gravity 
cases ($S^2$) given by Cho and Kantowski
\cite{chokan}. For matter
fluctuations only \cite{km2}, excluded are $N>14$ for a single
vector field and $N>6$ for a single tensor field.  (Fermions always have
a negative Casimir energy in even dimensions.)
Of course, it is possible to achieve cancellations by including
various matter fields and gravity. In general the Casimir energy is obtained
by summing over the species of field which propagate in the extra dimensions,
 \begin{equation}
 u_{\rm tot}={1\over a^4}
\sum_i\left[\alpha_i
\ln(a/L_{\rm Pl})+\beta_i\right]\approx{\beta_{\rm eff}\over a^4},
\end{equation}
which leads to a lower limit according to Eq.~(\ref{betalim}).
Presumably, if exact supersymmetry held in the extra dimensions (including
supersymmetric boundary conditions), the Casimir energy would vanish, but
this would seem to be difficult to achieve with {\em large\/} extra
dimensions (1 mm corresponds to $2\times 10^{-4}$ eV.)

It seems to be commonly believed that submillimeter tests of gravity
put no limits on the size of extra dimensions if $N>2$.  This is because
of the relation of the size $R$ of the extra dimensions in the ADD
scheme to the fundamental $4+N$ gravity scale $M$ \cite{addmore}:
\begin{equation}
R\sim{1\over M}\left(M_{\rm Pl}\over M\right)^{2/N},
\end{equation}
where $M_{\rm Pl}=1/L_{\rm Pl}=1.2\times 10^{16}$ TeV is the usual
Planck mass.  Moreover, the supernova limits on ADD extra dimensions
(due to production of Kaluza-Klein gravitons)
become rapidly smaller with increase in $N$ \cite{SNconstraints}:
\begin{mathletters}
\begin{eqnarray}
N=2:\quad R&<&0.9\times 10^{-4}\,\mbox{mm},\\
N=3:\quad R&<&1.9\times 10^{-7}\,\mbox{mm}.
\end{eqnarray}
\end{mathletters}  
Thus direct tests of Newton's law are not competitive.  However, the resulting
Casimir contribution to the cosmological constant would be enormous for
such small compactified regions, and it would seem impossible to naturally
resolve this problem.

The situation at first glance seems rather different with the RS scenario.  
In the original
scheme, gravity is localized in the ``Planck brane,'' while the standard-model
particles are confined to the ``TeV brane.''  As a consequence, it might appear
that the quantum fluctuations of both brane and bulk fields are negligible
\cite{goldroth}.  It has been stated that the cosmological constant becomes
exponentially small as the brane separation becomes large \cite{tye}.
However, this is at the ``classical level,'' without bulk fluctuations;
explicit considerations show that quantum effects give rise to a large
cosmological constant, of order of that given by Eq.~(\ref{ccprob}), 
unless an appeal is made to fine tuning \cite{odintsov}.
Moreover, if the scenario is extended so that the world
brane contains compactified dimensions in which gravity lives \cite{chako},
the constraints we deduce here directly apply.

There have been a number of proposals in which either in string 
\cite{rs,stringcc}
or brane \cite{branecc} contexts the cosmological constant can be made to
vanish.  These solutions may not be altogether natural, and may indeed
require fine tuning \cite{finetune}.  (Moreover, the suggestions either
do not include gravitational fluctuations, or ignore the problem
of gauge and parameterization dependence.) Such ideas could explain the
vanishing of $u_{\rm cosmo}$, or of a corresponding energy arising at the
electroweak symmetry breaking scale, $\Lambda_{\rm EW}\sim 1$ TeV, 
\begin{equation}
u_{\rm EW}\sim \Lambda_{\rm EW}^4\sim 10^{53} \mbox{GeV/cm}^3,
\end{equation}
due to fluctuations in standard model fields, but they would presumably not
lead to the simultaneous vanishing of the Casimir energy.  Indeed Sundrum
in Ref.~\cite{rs} obtains a small cosmological constant based on a narrow window
in allowed graviton compositeness, or string, scale, $m_{\rm st}$,
\begin{equation}
10\,\mu\mbox{m}<{1\over m_{\rm st}}<1 \,\mbox{cm}.
\end{equation}
Theoretical and experimental limits this past year have nearly closed this
window.  Clearly, the ideas expressed here provide a stringent constraint
for model builders.

%-----------------------------------------------------------------------
% THE ACKNOWLEDGEMENTS
%-----------------------------------------------------------------------
\section*{Acknowledgments}

This research was supported in part by the U.S. Department of Energy 
under Grant No.~DE-FG03-98ER41066 and by the National Science Foundation 
CAREER grant AST-0094335.

%-----------------------------------------------------------------------
%   REFERENCES
%-----------------------------------------------------------------------

\begin{table}
\begin{tabular}{lcccc}
Geometry $\cal S$&Gravity&Scalar&Fermion&Vector\\
\hline
$S^1$ (u)&$\beta=-2.53\times10^{-4}$&$-5.06\times10^{-5}$&$2.02\times10^{-4}$&
---\\
$S^1$ (t)&$\beta=2.37\times 10^{-4}$&$4.74\times 10^{-5}$&$-1.90\times10^{-4}$&
---\\
$S^2$&$\alpha=1.70\times10^{-2}$&$-8.04\times10^{-5}$&$-7.94\times10^{-4}$&
$-8.04\times 10^{-5}$\\
$S^3$&---&$\beta=7.57\times 10^{-5}$&$1.95\times10^{-4}$&---\\
$S^4$&$\alpha=-0.489$&$-4.99\times10^{-4}$&
$-6.64\times10^{-3}$&$1.21\times10^{-2}$\\
$S^5$&---&$\beta=4.28\times10^{-4}$&$-1.14\times10^{-4}$&---\\
$S^6$&$\alpha=5.10$&$-1.31\times 10^{-3}$&$-3.02\times10^{-2}$&$4.90\times
10^{-2}$\\
$S^7$&---&$\beta=8.16\times10^{-4}$&$5.96\times 10^{-5}$&---\\
\end{tabular}
\caption{The Casimir energy for $M^4\times {\cal S}$ is tabulated for various
field types in the compact geometry $\cal S$.  We write $u=[\alpha
\ln(a/L_{\rm Pl})+\beta]a^{-4}$, and give $\alpha$ for even internal dimension
and $\beta$ for odd, where $\alpha=0$. For $S^1$ u denotes untwisted 
(periodic) while t twisted (antiperiodic) boundary conditions.
 The entries marked with dashes
have not been calculated.  (We should note that a general recipe for calculating
the odd-$N$ terms for vectors and tensors
is given in Ref.~\protect\cite{chodosandmyers},
but results are not explicit, and require knowledge of the
``polylogarithmic-exponential'' function.  Moreover, the Casimir energies
they find are complex.  The method given in 
Ref.~\protect\cite{km2} has, in contrast,
no problems with tachyons, and would give real energies.
Explicit numbers given in Ref.~\protect\cite{sarmadi} for the
various components of gravity without the necessary Vilkovisky-DeWitt
correction do not allow one to extract the transverse vector part
without further calculation.)  The numbers are taken from 
Refs.~\protect\cite{cw,km,km2,chokan}.}
\label{tab1}
\end{table}

\begin{table}
\begin{tabular}{lcccc}
Geometry $\cal S$&Gravity&Scalar&Fermion&Vector\\
\hline
$S^1$ (u)&*&*&9.5 $\mu$m&---\\
$S^1$ (t)&9.9 $\mu$m&6.6 $\mu$m&*&---\\
$S^2$&84 $\mu$m&*&*&*\\
$S^3$&---&7.5 $\mu$m&9.5 $\mu$m&---\\
$S^4$&*&*&*&77 $\mu$m\\
$S^5$&---&11.5 $\mu$m&*&---\\
$S^6$&350 $\mu$m&*&*&110 $\mu$m\\
$S^7$&---&13.5 $\mu$m&7.0 $\mu$m&---\\
\end{tabular}
\caption{The lower limit to the radius of the compact dimensions
deduced from the requirement that the Casimir energy not exceed
the critical density. The numbers shown are for a single species of
the field type indicated.  The dashes indicate cases where the Casimir
energy has not been calculated, while asterisks indicate (phenomenologically
excluded) cases where the Casimir energy is negative.}
\label{tab2}
\end{table}

%-----------------------------------------------------------------------
%   END DOCUMENT
%-----------------------------------------------------------------------

\begin{references}
%
\bibitem{weinberg} S. Weinberg, Rev. Mod. Phys. {\bf 61}, 1 (1989);
Phys. Rev. D {\bf 61}, 103505 (2000); talk at Dark Matter 2000,
Marina del Rey, CA, 2000, astro-ph/0005265.
\bibitem{reiss} P.M. Garnavich {\it et al.},  Astrophys. J. Lett. 
{\bf 493}, L53 (1998);
A. G. Reiss {\it et al.}, Astron. J. {\bf 116}, 1009 (1998).
\bibitem{perlmutter} S. Perlmutter {\it et al.}, Astrophys. J. 
{\bf 517}, 565 (1999).
\bibitem{Boomerang}
The BOOMERANG collaboration:
P. de Bernardis  {\it et al.}, Nature, {\bf 404}, 955 (2000);
A.~E.~Lange {\it {\it et al.}},
%``Cosmological parameters from the first results of BOOMERANG,''
Phys.\ Rev.\ D {\bf 63}, 042001 (2001);
C.~B.~Netterfield {\it {\it et al.}},
%``A measurement by BOOMERANG of multiple peaks
%  in the angular power spectrum of the cosmic microwave background,''
astro-ph/0104460.
%\bibitem{Maxima}
%S. Hanany  {\it et al.}, astro-ph/0005123, Ap. J. Lett.,  in press. 

\bibitem{balbi}
The MAXIMA collaboration: 
A. Balbi {\it et al.}, Astrophys. J. Lett. {\bf 545}, L1 (2000);
%\bibitem{maxiboom} The Maxiboom Collaboration (J. Richard Bond {\it et al.})
%in Proceedings of {\it Conference on Cosmology and Particle Physics
%(CAPP 2000), Verbier, Switzerland, 17--28 July 2000}.
S.~Hanany {\it {\it et al.}},
% ``MAXIMA-1: A Measurement of
%   the Cosmic Microwave Background Anisotropy
%   on angular scales of 10 arcminutes to 5 degrees,''
Astrophys.\ J.\ Lett. {\bf 545}, L5 (2000).

\bibitem{DASI}
The DASI collaboration:
C.~Pryke {\it {\it et al.}},
%``Cosmological Parameter Extraction from the First Season
%  of Observations with DASI,''
astro-ph/0104490.
\bibitem{dielectricball} K. A. Milton and Y. J. Ng, Phys. Rev. E
{\bf 57}, 5504 (1998); I. Brevik, V. N. Marachevsky, and K. A. Milton,
Phys. Rev. Lett. {\bf 82}, 3948 (1999); G. Barton, J. Phys. A {\bf 32},
525 (1999); M. Bordag, K. Kirsten, and D. Vassilevich, Phys. Rev. D
{\bf 59}, 085011 (1999);
J. S. H\o ye and I. Brevik, J. Stat. Phys. {\bf 100}, 223 (2000).


\bibitem{kk} T. Kaluza, Sitz. Preuss. Akad. Wiss. Phys. Math. {\bf K1}, 966
(1921); O. Klein, Nature {\bf 118}, 516 (1926); Z. Phys. {\bf 37}, 895 (1926).
%
\bibitem{add} 
N.~Arkani-Hamed, S.~Dimopoulos and G.~Dvali,
%``The hierarchy problem and new dimensions at a millimeter,''
Phys.\ Lett.\ B {\bf 429}, 263 (1998); 
I.~Antoniadis, N.~Arkani-Hamed, S.~Dimopoulos and G.~Dvali,
%``New dimensions at a millimeter to a Fermi and superstrings at a TeV,''
Phys.\ Lett.\ B {\bf 436}, 257 (1998).
%
%\bibitem{dg} 
%S.~Dimopoulos and G.~F.~Giudice,
%``Macroscopic Forces from Supersymmetry,''
%Phys.\ Lett.\ B {\bf 379}, 105 (1996).
%
\bibitem{lrs} 
L.~Randall and R.~Sundrum,
%``A large mass hierarchy from a small extra dimension,''
Phys.\ Rev.\ Lett.\  {\bf 83}, 3370 (1999);
%``An alternative to compactification,''
Phys.\ Rev.\ Lett.\  {\bf 83}, 4690 (1999).
%
%\bibitem{add2} 
%N.~Arkani-Hamed, S.~Dimopoulos, G.~Dvali and N.~Kaloper,
%``Infinitely large new dimensions,''
%Phys.\ Rev.\ Lett.\  {\bf 84}, 586 (2000).
%
\bibitem{long} J. C. Long, H. W. Chan, and J. C. Price, Nucl. Phys. B 
{\bf 539}, 23 (1999); J. C. Long, A. B. Churnside, and J. C. Price, in
{\it Proceedings of the 9th Marcel Grossmann Meeting on Recent Developments
in Theoretical and Experimental General Relativity, Gravitation, and 
Relativistic Field Theories (MG9), Rome, 2--9 July 2000}, hep-ph/0009062.
\bibitem{hoyle} C. D. Hoyle, U. Schmidt, B. R. Heckel, E. G. Adelberger,
J. H. Gundlach, D. J. Kapner, and H. E. Swanson, Phys. Rev. Lett.
{\bf 86}, 1418 (2001).

\bibitem{most} M. Bordag, B. Geyer, G. L. Klimchitskaya, and V. M.
Mostepanenko, Phys. Rev. D {\bf 58}, 075003 (1998); {\bf 60}, 055004 (1999);
{\bf 62}, 011701(R) (2000); V. M. Mostepaneko and M. Novello, {\it ibid.}
{\bf 63}, 115003 (2001);
E. Fischbach, D.E. Krause, V.M. Mostepanenko, M. Novello,  hep-ph/0106331.
\bibitem{exp} S. K. Lamoreaux, Phys Rev. Lett. {\bf 78}, 5 (1997);
{\bf 81}, 5475(E) (1998);
U. Mohideen and A. Roy, Phys. Rev. Lett. {\bf 81}, 4549
(1998); {\bf 83}, 3341 (1999);
A. Roy, C.-Y. Lin, and U. Mohideen,
{\it Phys. Rev. D\/} {\bf 60}, R111101 (1999);
B. W. Harris, F. Chen, and U. Mohideen, {\it Phys. Rev. A\/}
{\bf 62}, 052109 (2000);
H. B. Chan, V. A. Aksyuk, R. N. Kleiman, D. J. Bishop, and F. Capasso,
Science {\bf 291}, 1941 (2001) (10.1126/science.1057984).

\bibitem{ac} T. Appelquist and A. Chodos, Phys. Rev. Lett. {\bf 50}, 141
(1983); Phys. Rev. D {\bf 28}, 772 (1983). 
\bibitem{cw} P Candelas and S. Weinberg, Nucl. Phys. B {\bf 237}, 397 (1984).
\bibitem{km} R. Kantowski and K. A. Milton, Phys. Rev. D {\bf 35}, 549 (1987).
\bibitem{km2} R. Kantowski and K. A. Milton, Phys. Rev. D {\bf 36}, 3712
(1987).
\bibitem{chodosandmyers} A. Chodos and E. Myers, Phys. Rev. D {\bf 31}, 3064
(1985).
\bibitem{sarmadi} M. H. Sarmadi, Nucl. Phys. B {\bf 263}, 187 (1986).
\bibitem{birm} D. Birmingham, R. Kantowski, and K. A. Milton, Phys. Rev. D
{\bf 38}, 1809 (1988).

\bibitem{kaplanwise} D. B. Kaplan and M. B. Wise, JHEP {\bf 08}, 037 (2000);
T. Banks, Nucl. Phys. B {\bf 309}, 493 (1988); S. R. Beane, Phys. Lett. B
{\bf 358}, 203 (1995); Gen. Rel. Grav. {\bf 29}, 945 (1997).

\bibitem{chen} J.-W. Chen, M. A. Luty, and E. Pont\'on, JHEP {\bf 09},
012 (2000).
 
\bibitem{devil} G. A. Vilkovisky, in {\it Quantum Theory of Gravity}, ed.\
S. C. Christensen (Hilger, Bristol, 1984); Nucl. Phys. B {\bf 234}, 125 (1984);
A. O. Barvinsky and G. A. Vilkovisky, Phys. Rep. {\bf 119}, 1 (1985);
B. S. DeWitt, in {\it Quantum Field Theory and Quantum Statistics}, ed.\
I. A. Batalin, C. J. Isham, and G. A. Vilkovisky (Hilger, Bristol, 1987).
\bibitem{ck91} H. T. Cho and R. Kantowski, Phys. Rev. Lett. {\bf 67}, 422 
(1991); {\bf 75}, 2630(E) (1995).
\bibitem{chokan} H. T. Cho and R. Kantowski, Phys. Rev. D {\bf 62}, 124003
(2000).
\bibitem{pdg} Particle Data Group, Eur. Phys. J. C {\bf 15}, 1 (2000);
W. J. Freedman {\it et al.}, astro-ph/0012376, Ap. J., to be published;
D. Branch, {\it Annu. Rev.  Astron. Astrophys.} {\bf 36}, 17 (1998), 
astro-ph/9801065; and references therein.
\bibitem{addmore} N. Arkani-Hamed, S. Dimopoulos, and G. Divali,
Phys. Rev. D {\bf 59}, 086004 (1999);
G. F. Guidice, R. Rattazzi, and J. D. Wells, Nucl. Phys. B {\bf 544} 3 (1999);
E. A. Mirabelli, M. Perelstein, and M. E. Peshkin, Phys. Rev. Lett. {\bf 82},
2236 (1999).
\bibitem{SNconstraints} 
S.~Hannestad and G.~Raffelt, 
%``New supernova limit on large extra dimensions,'' 
hep-ph/0103201; 
V.~Barger, T.~Han, C.~Kao and R.~J.~Zhang,
%``Astrophysical constraints on large extra dimensions,''
Phys.\ Lett.\ B {\bf 461}, 34 (1999);
S.~Cullen and M.~Perelstein,
%``SN1987A constraints on large compact dimensions,''
Phys.\ Rev.\ Lett.\  {\bf 83}, 268 (1999).

\bibitem{goldroth} W. D. Goldberger and I. Z. Rothstein, Phys. Lett. B 
{\bf 475}, 275 (2000); I. Brevik, K. A. Milton, S. Nojiri, and S. D. Odintsov,
Nucl. Phys. B {\bf 599}, 305 (2001).
\bibitem{tye} S.-H. H. Tye and I. Wasserman, Phys. Rev. Lett. {\bf 86}, 1682
(2001).
\bibitem{odintsov} S. Nojiri, O. Obregon, and S. D. Odintsov, hep-th/0105300.
\bibitem{chako} Z. Chako, P. J. Fox, A. E. Nelson, and N. Weiner,
hep-ph/0106343.

\bibitem{rs} R. Sundrum, JHEP {\bf 9907}, 001 (1999).
\bibitem{stringcc} 
 S. Kachru and E. Silverstein, JHEP {\bf 01}, 004 (1999);
S. Kachru, J. Kumar, and E. Silverstein, Phys. Rev. D {\bf 59}, 106004 (1999);
S. Kachru, M. Schulz, and E. Silverstein, Phys. Rev. D {\bf 62}, 045021 (2000).
\bibitem{branecc} N. Arkani-Hamed, S. Dimopoulos, N. Kaloper, and R. Sundrum,
Phys. Lett. B {\bf 480}, 193 (2000); S. F\"orste, Z. Lalak, S. Lavignac,
and H. P. Nilles, Phys. Lett. B {\bf 481}, 360 (2000).  
\bibitem{finetune} S. F\"orste, Z. Lalak, S. Lavignac, and H. P. Nilles, 
JHEP {\bf 09}, 034 (2000).
\end{references}
\end{document}